\documentclass[superscriptaddress,showkeys,twocolumn,nofootinbib,longbibliography]{revtex4-2}
\def\letter{1}\def\pr{1}
\usepackage{epsfig}
\usepackage{epstopdf}
\usepackage{graphicx}
\usepackage{ifthen}
\usepackage{ulem}

\usepackage{float}

\usepackage{amsmath}
\usepackage[psamsfonts]{amssymb}
\usepackage{euscript}

\usepackage{latexsym}
\usepackage[arrow,matrix,curve]{xy}

\usepackage[hypertexnames=false]{hyperref}
\hypersetup{
    colorlinks=true,
    linkcolor=blue,
    filecolor=magenta,   
    citecolor=black,   
    urlcolor=cyan,
    }

\newskip\humongous \humongous=0pt plus 1000pt minus 1000pt

\newif\ifdtup

\def\,{\hspace{-.1cm}}
\def\hsp{,\hspace{.7cm}}

\def\fc#1#2 {\frac{n}{q}#1\frac{n}{q}#2}

\newcommand{\vac}{\ensuremath{|0\rangle}}

\def\dft{\mathcal{D}_{f}^{(t)}}

\renewcommand{\cos}{\textrm{cos}}
\renewcommand{\sin}{\textrm{sin}}

\renewcommand{\tanh}{\textrm{tanh}}
\newcommand{\sech}{\textrm{sech}}

\def\exp#1{\hbox{\rm exp}\left(#1\right)}

\renewcommand{\theequation}{\arabic{section}.\arabic{equation}}
\renewcommand{\(}{\begin{equation}}
\renewcommand{\)}{end{equation} \vspace{-.05in}\linebreak}

\newcounter{saveeqn}
\newcounter{savealpheqn}

\newcommand{\alpheqn}{\setcounter{saveeqn}{\value{equation}}%
  \stepcounter{saveeqn}\setcounter{equation}{0}%
  \renewcommand{\theequation}{\mbox{\arabic{section}.\arabic{saveeqn}
\alph{equation}}}
  \renewcommand{\)}{\end{equation}}}
\def\part#1{\frac{\partial}{\partial{#1}}}%
\def\group#1{\refstepcounter{equation}\setcounter{saveeqn}
 {\value{equation}}%
  \label{#1}\setcounter{equation}{0}%
\renewcommand{\theequation}{\mbox{\arabic{section}.\arabic{saveeqn}
\alph{equation}}}
  \renewcommand{\)}{\end{equation}}}
\newcommand{\reseteqn}{\setcounter{equation}{\value{saveeqn}}%
  \renewcommand{\theequation}{\arabic{section}.\arabic{equation}}%
  \renewcommand{\)}{\end{equation}}}

\newcommand{\aalpheqn}{\setcounter{saveeqn}{\value{equation}}%
  \stepcounter{saveeqn}\setcounter{equation}{0}%
  \renewcommand{\theequation}{\mbox{
        \Alph{subsection}.\arabic{saveeqn}\alph{equation}}}
   \renewcommand{\)}{\end{equation}}}
\newcommand{\areseteqn}{\setcounter{equation}{\value{saveeqn}}%
  \renewcommand{\theequation}{\Alph{subsection}.\arabic{equation}}%
  \renewcommand{\)}{\end{equation}}}

\renewcommand{\thefootnote}{\alph{footnote}}
\renewcommand{\(}{\begin{equation}}
\renewcommand{\)}{\end{equation}}
\newcommand{\ba}{\begin{eqnarray}}
\newcommand{\ea}{\end{eqnarray}}

\renewcommand{\sl}{{g}}

\newcommand{\cbp}{\mathop{\vtop{\ialign{##\crcr
   $\hfil\displaystyle{}\hfil$\crcr\noalign{\kern-13pt\nointerlineskip}
   \BIG{)}\hskip0pt\crcr\noalign{\kern3pt}}}}}
\newcommand{\pa}{\mathop{\vtop{\ialign{##\crcr

$\hfil\displaystyle{\oplus}\hfil$\crcr\noalign{\kern+1pt\nointerlineskip
}
   \hspace{.08in}$^{\alpha=0}$\hskip6pt\crcr\noalign{\kern3pt}}}}}
\renewcommand{\hsp}{,\hspace{.3in}}
\newcommand{\p}{^\prime}
\newcommand{\pp}{^{\prime\prime}}

\catcode`\@=11
\def\vereq#1#2{\lower3pt\vbox{\baselineskip1.5pt \lineskip1.5pt
\ialign{$\m@th#1\hfill##\hfil$\crcr#2\crcr\sim\crcr}}}
\catcode`\@=12

\renewcommand{\(}{\begin{equation}}
\renewcommand{\)}{\end{equation}}

\def\pin#1{\int \frac{d#1}{2\pi}}

\def\df{\mathcal{D}_{f}}

\def\cG{{\mathcal{G}}}
\def\cH{{\mathcal{H}}}

\newcommand{\beas}{\begin{eqnarray*}}
\newcommand{\eeas}{\end{eqnarray*}}

\newcommand{\bquo}{\begin{quote}}
\newcommand{\enqu}{\end{quote}}

\def\lim#1{\stackrel{\rm{lim}}{{}_{#1}}}

\newcommand{\g}{\mathfrak g}

\def\ch{{\mathcal{H}}}

\def\ok#1{\omega_{k_{#1}}}

\newcommand{\beq}{\begin{equation}}
\newcommand{\eeq}{\end{equation}}
\newcommand{\bea}{\begin{eqnarray}}
\newcommand{\eea}{\end{eqnarray}}

\newskip\humongous \humongous=0pt plus 1000pt minus 1000pt

\newif\ifdtup

\catcode`\@=11

\ifthenelse{\equal{\letter}{0}}{ 

\@addtoreset{equation}{section}
\def\theequation{\arabic{section}.\arabic{equation}}

\def\@normalsize{\@setsize\normalsize{15pt}\xiipt\@xiipt
\abovedisplayskip 14pt plus3pt minus3pt%
\belowdisplayskip \abovedisplayskip
\abovedisplayshortskip \z@ plus3pt%
\belowdisplayshortskip 7pt plus3.5pt minus0pt}

\def\small{\@setsize\small{13.6pt}\xipt\@xipt
\abovedisplayskip 13pt plus3pt minus3pt%
\belowdisplayskip \abovedisplayskip
\abovedisplayshortskip \z@ plus3pt%
\belowdisplayshortskip 7pt plus3.5pt minus0pt
\def\@listi{\parsep 4.5pt plus 2pt minus 1pt
      \itemsep \parsep
      \topsep 9pt plus 3pt minus 3pt}}

\relax

\def\section{\@startsection{section}{1}{\z@}{3.5ex plus 1ex minus  .2ex}{2.3ex plus .2ex}{\large\bf}}

\def\thesection{\arabic{section}}
\def\thesubsection{\arabic{section}.\arabic{subsection}}

\def\appendix{\setcounter{section}{0}
 \def\thesection{Appendix \Alph{section}}
 \def\thesubsection{\Alph{section}.\arabic{subsection}}
 \def\theequation{\Alph{section}.\arabic{equation}}}
\renewcommand{\theequation}{\arabic{section}.\arabic{equation}}

}{
\renewcommand{\theequation}{\arabic{equation}}

} 

\def\appendix{\setcounter{section}{0}
 \def\thesection{Appendix \Alph{section}}
 \def\thesubsection{\Alph{section}.\arabic{subsection}}
 \def\theequation{\Alph{section}.\arabic{equation}}}

\begin{document}
\def\thefootnote{\fnsymbol{footnote}}
\def\thetitle{Quantum Oscillons are Long-Lived}
\def\auttwo{Katarzyna Slawi\'nska}
\def\autone{Jarah Evslin}
\def\autthree{Tomasz Roma\'nczukiewicz}
\def\autfour{Andrzej Wereszczy\'nski}

\def\affc{Institute  of  Theoretical Physics,  Jagiellonian  University,  Lojasiewicza  11,  Krak\'ow,  Poland}

\def\affd{International Institute for Sustainability with Knotted Chiral Meta Matter (WPI-SKCM$^{\; 2}$), Hiroshima University, 1-3-1 Kagamiyama, Higashi-Hiroshima, Hiroshima 739-8531, JAPAN}

\def\affb{University of the Chinese Academy of Sciences, YuQuanLu 19A, Beijing 100049, China}
\def\affa{Institute of Modern Physics, NanChangLu 509, Lanzhou 730000, China}


\ifthenelse{\equal{\pr}{1}}{
\title{\thetitle}
\author{\autone}
\affiliation {\affa}
\affiliation {\affb}
\author{\auttwo}
\author{\autthree}
\affiliation {\affc}
\author{\autfour}
\affiliation {\affc}
\affiliation {\affd}

}{

\begin{center}
{\large {\bf \thetitle}}

\bigskip

\bigskip


{\large \noindent  \autone{${}^{1,2}$} \footnote{jarah@impcas.ac.cn} {, \auttwo{${}^{3}$} \footnote{katarzyna.slawinska@uj.edu.pl},
\autthree{${}^{3}$} \footnote{tomasz.romanczukiewicz@uj.edu.pl},
and \autfour{${}^{3}$} \footnote{andrzej.wereszczynski@uj.edu.pl}
}}


\vskip.7cm

1) \affa\\
2) \affb\\
3) \affc\\

\end{center}

}

\begin{abstract}
\noindent
As the longest lived transient, oscillons play a critical role in classical field theory simulations of many phenomena.  However, beyond the classical approximation, it is well-known that quantum corrections open decay channels through which oscillons radiate rapidly.  Therefore it is believed that in the real world, oscillons are too short-lived to be phenomenologically relevant.  We observe that previous calculations of the radiated power assume that the oscillon is in a coherent state.  We show that a squeezed coherent state, on the other hand, would emit no radiation at leading order in the coupling.  This leads us to the conclusion that the instantaneous radiation calculated in the literature corresponds not to the oscillon's decay, but rather to its relaxation from a coherent state to a lower-energy, squeezed coherent state, which then radiates much more slowly.  As a result, the lifetime of the quantum oscillon is enhanced by an inverse power of the coupling.

\end{abstract}

%
\setcounter{footnote}{0}
\renewcommand{\thefootnote}{\arabic{footnote}}

\ifthenelse{\equal{\pr}{1}}
{
\maketitle
}{}
\section{Introduction}
Oscillons are the most generic soliton-like excitations.  In contrast with topological solitons, oscillons decay in classical field theory \cite{osc76}.  However, they decay extraordinarily slowly \cite{gleiser94, Hind, oriol}.  For example, in 1+1 dimensions, the small-amplitude oscillons with widths of order $1/\epsilon$, have lifetimes of order $e^{-1/\epsilon.}$  This long lifetime, together with the fact that they exist in many different models, means that oscillons dominate the energy budgets at intermediate times after many violent events  \cite{G-cosm,oscrev}.  As a result, simulations suggest that they play an important role in phenomena from bubble collapse \cite{osc1,decosc}, to reheating \cite{osc2,re2} to soliton collisions \cite{vort, osc3,scat,CS,CS-we} and perhaps even the formation of primordial black holes \cite{bhosc}. This makes them relevant for the evolution of the early Universe \cite{Aur} and a candidate for dark matter \cite{Olle, Kaw}. Oscillons may also potentially exist in fundamental theories, see {\it{e.g.}} ~\cite{Gr}. 

It thus came as a surprise in Refs.~\cite{hertz,vach} when it was discovered that, beyond the classical approximation, the leading quantum corrections cause oscillons to decay {\it very rapidly}.  The oscillon lifetime was approximated to be the oscillon's energy divided by the instantaneous power radiated, yielding a lifetime so short that in much of parameter space the oscillons became phenomenologically irrelevant.

While those calculations were correct, we will show that the oscillon state considered in those calculations was a coherent state.  Even in the better understood case of the quantum kink, it is known that the true ground state is a squeezed coherent state \cite{cahill76,mekink}, which has a lower energy than the classical soliton \cite{dhn2} whose energy is equal to that of the coherent state.  The coherent state is thus an excited state, and it was shown in Ref.~\cite{noiosc} that the quantum kink coherent state radiates a burst of energy just like that observed for the oscillon, but that this radiation turns off after a time of order $O(1/m)$, where $m$ is the mass of a fundamental meson, once the kink has relaxed to a squeezed coherent state.

In this letter we will argue that the same story applies to the oscillon.  We will construct the corresponding squeezed coherent oscillon state and will show that, at leading order in the coupling and the oscillon thickness, the state is periodic.  We will also argue that Ref.~\cite{hertz} assumed the oscillon to be in a coherent state\footnote{Ref.~\cite{vach} uses an adiabatic approximation which is also equivalent to the choice of a coherent state.}.  We then conclude that the high instantaneous power radiated by the quantum oscillon in Refs.~\cite{hertz,vach} will stop once the oscillon state is squeezed, and so it does not represent an instability of the quantum oscillon. 

\section{The Oscillon}

We will consider a 1+1 dimensional model of a Schrodinger picture scalar field $\phi(x)$ with conjugate momentum $\pi(x)$, described by the Hamiltonian
\beq
H=\int dx :\ch(x):
\eeq
where the Hamiltonian density is
\beq
\ch(x)=\frac{\pi(x)^2+\partial_x\phi(x)\partial_x\phi(x)}{2}+\frac{V(g\phi(x))}{g^2}.
\eeq
We consider a general potential $V$, but we will assume that it has a local or global minimum and will choose conventions so that
\beq
V\p(0)=V(0)=0.
\eeq
Here $g$ is the coupling constant and we define the scalar's mass squared to be
\beq
m^2=V\pp(0) >0.
\eeq

Letting $V^{(n)}$ be the $n$th derivative of the potential at zero, we can define the coupling
\beq
\lambda_F=g^2\left(\frac{5V^{(3)2}}{6m^2}-\frac{V^{(4)}}{2}\right).
\eeq
When it is positive, this theory admits standard oscillon solutions. We remark that this is not a necessary condition and exotic oscillons violating $\lambda_F>0$ are known \cite{Dor, Blaschke, vD}. 

We will be interested in {\it small amplitude}, that is, {\it thick} oscillons, with a spatial width of $1/\epsilon$.  These can be constructed \cite{fodor} order by order in $\epsilon/m$, beginning with
\beq
f(x,t)=2\sqrt\frac{2}{\lambda_F}{\rm{sech}}\left( 
\epsilon x
\right) {\rm{cos}}(\Omega t),\ \  \Omega=\sqrt{m^2-\epsilon^2}.
\eeq
At every order in $\epsilon$, the oscillon is periodic with frequency smaller than the mass threshold, which prohibits linear coupling to radiation. The classical oscillon's decay rate is slower than any power of $\epsilon$.  In this note we will work at the leading order in $\epsilon$, and so the classical oscillon is approximated to be periodic.

\section{Linearized Perturbations}

At the leading order in $\epsilon$, the linearized perturbations of the small-amplitude oscillon include four discrete modes ~\cite{oscnormal}
\bea
\g_B(\epsilon x,t)&=& 
{\rm{tanh}}\left(
\epsilon x
\right) {\rm{sech}}\left( 
\epsilon x
\right)\cos\left( \Omega t\right)\nonumber\\
\g_T(\epsilon x,t)&=&
{\rm{sech}}\left( 
\epsilon x
\right)\sin\left( \Omega t\right)\nonumber\\
\g_M(\epsilon x,t)&=&t\g_B(\epsilon x,t)+x\g_T(\epsilon x,t)\nonumber\\
\g_{\epsilon}(\epsilon x,t)&=&
{\rm{sech}}\left(
\epsilon x
\right)\cos\left( \Omega t\right)
\eea
which respectively describe spatial translations, time translations, boosts and increases in the oscillation amplitude.  There are also continuum Floquet modes
$\g(x,t+2\pi/\Omega)=e^{-i 2\pi \omega/\Omega}\g(x,t)$:
\beq
\g_k(\epsilon x,t)=\cG_k(\epsilon x)e^{-i(\Omega+\omega_k) t} +\cH_k(\epsilon x)e^{i(\Omega-\omega_k) t}\label{g1eq}
\eeq
where
\bea
\cG_k(\epsilon x)\,&=&\,\left(\frac{k^2}{m^2}\,-\,\tanh^2(\epsilon x)\,-\,\frac{2ik}{m}\tanh(\epsilon x)\right)e^{-i\epsilon x k/m},\nonumber\\
\mathcal{H}_k(\epsilon x)\,&=&\,\sech^2(\epsilon x)e^{-i\epsilon x k/m},\label{fl} 
\eea
and
\beq
\omega_k=\sqrt{m^2+\epsilon^2k^2/m^2}-\Omega,
\eeq
\begin{figure}[H]
\begin{center}
\includegraphics[width=1.6in,height=1.2in]{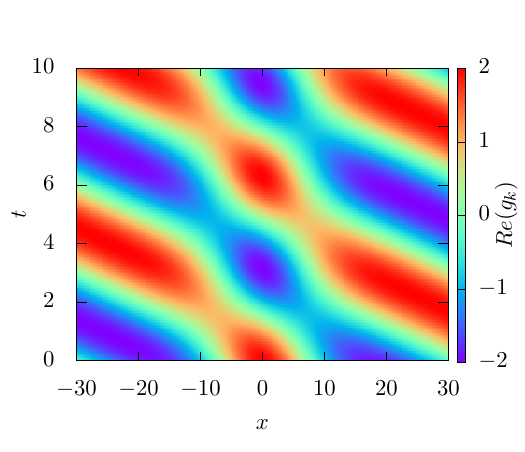}
\includegraphics[width=1.6in,height=1.2in]{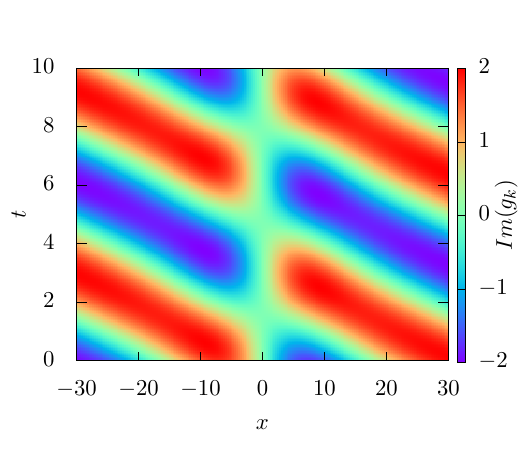}

\includegraphics[width=1.6in,height=1.2in]{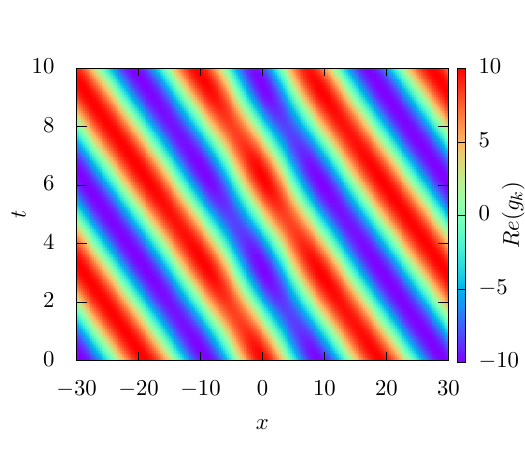}
\includegraphics[width=1.6in,height=1.2in]{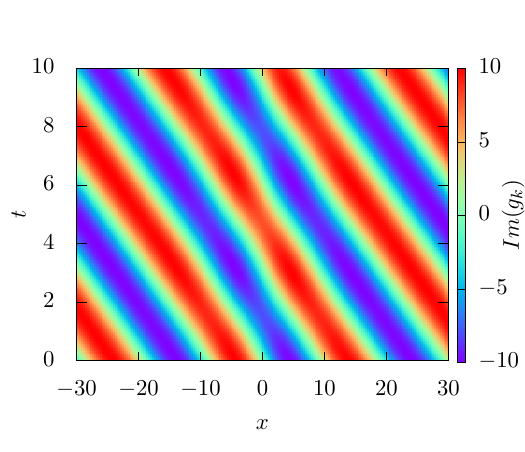}
\caption{The real (left) and imaginary (right) parts of the long wavelength continuum modes $\g_k(\epsilon x,t)$ at $k=m$ (top) and $k=3m$ (bottom).  Here $\epsilon=0.1$.  When $k/m\gg 1$ so that the wavelength is much shorter than the oscillon, these tend to plane waves.}
\label{contfig}
\end{center}
\end{figure}
\noindent with wavenumber $\epsilon k/m$. In Fig. \ref{contfig} we plot some of the continuum modes. Multiplying by a normalization factor of $m/k$ one sees that, when $k\gg m$ so that the modes are much thinner than the thick, small-amplitude oscillon, these reduce to plane waves.  On the other hand, when $k\sim m$ so that the length of the modes is of order the length of the oscillon, $\omega_k$ is of order $O(\epsilon^2)$.

\section{Decomposing the Field}

In Schrodinger picture quantum field theory, one may decompose the field in any basis of functions.  In the case of the quantum kink, Ref.~\cite{cahill76} decomposed $\phi(x)$ and $\pi(x)$ in terms of the kink's normal modes to diagonalize the free part of the Hamiltonian.  We will now perform a similar trick for the oscillon.  However, as the oscillon is time-dependent, instead of normal modes one has Floquet modes \cite{pars}.

Our proposed decomposition is
\bea
\phi(x)&=&\phi_0
\g_{B}(x,0)+\hat\epsilon\g_\epsilon(x,0)\label{decd}\\
&&+\pin{k}\left(\mathcal{G}_k(\epsilon x)+ \mathcal{H}_k(\epsilon x)
\right)\left(b_{-k}  +b^\dag_k \right)\nonumber
\\
\pi(x)&=&\phi_T\dot\g_T(x,0)+\pi_0 \dot\g_M(x,0)\nonumber\\
&&\ \ +i\pin{k}\left[\Omega\left(\mathcal{G}_k(\epsilon x)- \mathcal{H}_k(\epsilon x)\right)\right.\nonumber\\
&&\left.+\omega_k\left(\mathcal{G}_k(\epsilon x)+ \mathcal{H}_k(\epsilon x)\right)
\right]\left(b_{-k}-b^\dag_k\right). \nonumber 
\eea
The above identifications of the perturbation modes can be used to intuitively understand the operators introduced in this decomposition.  $\phi_0$ is the position of the oscillon. $\hat\epsilon$ is its amplitude.  $\phi_T$ is the temporal phase of the oscillon in its periodic evolution.  $\pi_0$ is the oscillon's velocity.  The operators $b^\dag_k$ and $b_k$ create and annihilate the Floquet mode with wavenumber $\epsilon k/m$.  The four discrete operators $\phi_0$, $\phi_T$, $\hat\epsilon$ and $\pi_0$ are not collective coordinates, but they are proportional to collective coordinates at leading order in perturbation theory as shown in \ref{ccapp}.

The decomposition of the field can be inverted to obtain the operators as described in \ref{flapp}.  We will need
\bea
\pi_0&=&\frac{\epsilon^2}{\Omega^2}\int dx\ \pi(x) \ \tanh{(\epsilon x)}\ \sech{(\epsilon x)}\label{pi0}\\
\hat{\epsilon}&=&\frac{\epsilon}{2}\int dx\ \phi(x)\  \sech{(\epsilon x)}\nonumber\\
b_k&=&\frac{1}{2C_k}\int dx \left[\left(\cG_{-k}(\epsilon x)-\cH_{-k}(\epsilon x)\right.\right.\nonumber\\
&&\left.+
\frac{\pi\ok{}^2}{2m^2}\ \sech\left(\frac{k\pi}{2m}
\right)\sech(\epsilon x)\right)\phi(x)\nonumber\\
&&-\frac{i}{\Omega} \left(\cG_{-k}(\epsilon x)+\cH_{-k}(\epsilon x)\right.\nonumber\\
&&\left.\left.\hspace{-.5cm}-i
 \frac{\epsilon\pi k \ok{}^2}{3m^2\Omega^2}\sech\left(\frac{k\pi}{2m}
\right)
\tanh(\epsilon x)\sech(\epsilon x)\right)\pi(x)
\right].\nonumber
\eea

Imposing that $\phi(x)$ and $\pi(x)$ satisfy canonical commutation relations, $[\phi(x),\pi(y)]=i\delta(x-y)$, one finds that $\pi_0$, $\hat\epsilon$ and $b_k$ all commute
\beq
[\pi_0,\hat{\epsilon}]= [\pi_0,b_k]=[\hat{\epsilon},b_k]=[b_{k_1},b_{k_2}]=0.
\eeq
These commutators will be critical in what follows, as our proposed oscillon state will be a simultaneous eigenstate of $\pi_0$, $\hat{\epsilon}$ and $b_k$.

\section{The Displacement Operator}

Configurations in classical field theory correspond to states in quantum field theory \cite{skyrme}.  To what state does the oscillon correspond?  Intuition from quantum solitons \cite{vinc72,cornwall74} suggests that it should be something like a coherent state, which is constructed by acting the displacement operator
\beq
\dft=\exp{-i\int dx \left(f(x,t)\pi(x)-\partial_t f(x,t) \phi(x)\right)}
\eeq
on the tree level vacuum $|\Omega\rangle$.  However, even in the case of quantum solitons, this is not quite the right state because it has the same energy as the classical solution, whereas quantum solitons have less energy \cite{dhn2}.  The solution, in that case, is that $|\Omega\rangle$ needs \cite{cahill76,mekink} to be replaced with a squeezed state $\vac$, where the squeeze is chosen so as to place all normal modes in their ground state.  

We will now attempt to replicate this strategy for the oscillon.  Let us consider a candidate oscillon state $|B\rangle_0$ and define $\vac_0$ as whatever state you obtain when you peel off the displacement operator
\beq
\vac_0=\df^{(t)\dagger}|B\rangle_0.
\eeq

\section{Time Evolution}

The time evolution of $|B\rangle_0$, like any state, is generated by the Hamiltonian $H$.  This means that the time evolution of the ket $\vac_0$, defined to be $\df^{(t)\dagger}$ of the evolved state $|B\rangle$, is generated by \cite{noiosc}
\beq
H\p=\df^{(t)\dag}H\dft +i\df^{(t)\dag}[\dft,\partial_t].
\eeq
Up to an irrelevant $c$-number, this is equal to
\beq
H\p_2\,=\,\frac{1}{2}\int dx\,\left(\pi^2(x)\,+\,(\partial_x\phi(x))^2\,+\,V^{(2)}(\sl f(x,t))\phi^2(x)\right) \label{hpe}
\eeq
plus interaction terms which are suppressed by powers of the coupling $g$.  In the rest of this letter we will ignore these interactions, and so our analysis will not be reliable at timescales of order $O(1/g)$.   

Note that this approximation is not the same as ignoring all interactions in the original potential $V$.  For example, if $V$ is a quintic, then $H\p_2$ would have a mass term with a mass squared of $g^3f^3$, and so one would find the equation of motion reported in Eq.~(41) of Ref.~\cite{hertz} which is responsible for the radiation observed in that study.  Therefore, by truncating $H\p$ at this order, we are not removing the radiation observed in Ref.~\cite{hertz}, which is not suppressed by any powers of the coupling. 

The time evolution can be computed using the time-ordered evolution operator
\beq
U(t)=T{\rm{exp}}\left(-i\int_0^t d\tau H\p_2(\tau)  
\right) \label{ut}
\eeq
together with the commutators
\beq
[H\p_2(t),\phi(x)]=-i\pi(x)
\eeq
and
\beq
[H\p_2(t),\pi(x)]
=-i\partial^2_x\phi(x)+iV^{(2)}(gf(x,t))\phi(x).
\eeq
In \ref{evapp} we find
\bea
U^\dag(t)\phi(x)U (t)&=&
\phi_0
\g_{B}(x,t)+\hat\epsilon\g_\epsilon(x,t)\label{pev}\\
&&+\pin{k}\g_{k}(x,t) \left(b_{-k}  +b^*_k \right)
\nonumber\\
U^\dag(t)\pi(x)U (t)&=&\phi_T\dot\g_T(x,t)+\pi_0 \dot\g_M(x,t)\nonumber\label{eveq}\\
&&+\pin{k}\dot{\g}_{k}(x,t)\left(- b_{-k}+b^*_k\right).\nonumber
\eea
Inserting $t=2\pi/\Omega$ and using the forms of the functions $\g$, one finds that the evolution of the field by a period leaves it invariant up to a $k$-dependent phase in the continuum terms.

One can then read off the evolution of the operators
\bea
U^\dag (2\pi/\Omega)\pi_0U (2\pi/\Omega)&=&\pi_0\label{per}\\
U^\dag(2\pi/\Omega)\hat{\epsilon}U (2\pi/\Omega)&=&\hat\epsilon\nonumber\\
U^\dag(2\pi/\Omega)b_kU(2\pi/\Omega)&=&e^{-2\pi i\ok{}/\Omega} b_k.\nonumber
\eea

\section{The Oscillon State}

Finally we are ready to present our squeezed state $\vac_0$.  It is the state such that
\beq
\pi_0\vac_0=b_k\vac_0=0\hsp \hat\epsilon\vac_0=\epsilon\vac_0
\eeq
for some oscillation amplitude $\epsilon>0$.  These conditions are consistent because $\pi_0$, $b_k$\ and $\hat{\epsilon}$ commute.

The periodicity conditions (\ref{per}) imply that the same conditions are satisfied after one period $2\pi/\Omega$, and so the squeezed state $\vac_0$ comes back to itself.

What about the oscillon state $|B\rangle_0=\dft\vac_0$?  The displacement operator $\dft$ is also periodic with period $2\pi/\Omega$, and so since $\vac_0$ is periodic, we conclude that $|B\rangle_0$ is itself periodic.

We have then shown that, up to corrections suppressed by powers of $\epsilon$ and the coupling $g$, our oscillon state $|B\rangle_0$ is periodic.  It therefore does not decay on timescales at which these approximations apply, in particular excluding the $O(m^2)$ radiated power in Ref.~\cite{hertz}.

The above discussion is only phenomenonologically relevant if $|B\rangle_0$ is an attractor, so that the instability leads the oscillon to this state.  In the case of the kink, this was shown in Ref.~\cite{noiosc}.  The argument here is quite similar.  The unsqueezed state may be created from the squeezed state by acting with a squeeze operator, which is an exponential in a bilinear in the modes.  This creates pairs of excitations on top of the state $|B\rangle_0$.  In the case in which both excitations are continuum Floquet modes, the excitations simply fly away at the speeds of the corresponding modes in the vacuum.  The coherent state is spatially localized while the squeezed state corresponds to a flat position-space wave function of oscillons.  Correspondingly, in terms in which one or both of the modes are translation zero modes, the system will relax via the usual wave packet spreading.   While the coherent state has a fixed temporal phase, our state $|B\rangle_0$ is a flat superposition of all phases. The evolution of the time zero modes is thus more interesting and, paralleling the evolution of a coherent state in the quantum harmonic oscillator, it remains coherent at leading order, rather than radiating.  

\vspace*{0.4cm}

In order to understand the above results on the quantization of oscillons one should start with a very well known case of the kink. The ground state of the quantum kink is a squeezed coherent state.  If one considers initial conditions which are instead a coherent state which is not squeezed, corresponding to acting the displacement operator on the free vacuum $|\Omega\rangle$, then one obtains an excited kink.  In Ref.~\cite{noiosc} it was shown that this excited kink relaxes via the emission of radiation with power of order $O(m^2)$, until the state has been squeezed.

In the case of the oscillon, Ref.~\cite{hertz} has implicitly used a coherent but not squeezed state $\dft|\Omega\rangle$.  Instead of acting $\dft$ on the vacuum, the fields were conjugated by $\dft$, yielding the shift reported in Eq.~(25) of that paper.  Conjugating the fields by $\dft$ and acting $\dft$ on the state are equivalent operations, in the sense that they yield the same matrix elements.  After this conjugation, the paper used the free vacuum $|\Omega\rangle$ for calculations.  As a result, the calculation assumed that the oscillon state is a coherent state $\dft|\Omega\rangle$.  

The radiation observed in that paper is then simply a consequence of the fact that $|\Omega\rangle$ is not an eigenstate of $H\p_2$.  Our $\vac_0$ state is, on the other hand, constructed so as to be periodic under the action of $H\p_2$.  We therefore interpret the radiation observed in that paper as the relaxation of the oscillon from $\dft|\Omega\rangle$ to $\dft\vac_0$.  We expect that, once the state has approached $\dft\vac_0$ and so is squeezed, the radiation will slow significantly.  As a result, the approximation, in Ref.~\cite{hertz}, that the oscillon lifetime is equal to the oscillon energy divided by the instantaneous radiated power will fail.

\vspace*{0.4cm}

We have not yet calculated all of the commutators of the operators in our decomposition (\ref{decd}).  If $\phi_T$ and $\hat\epsilon$ do not commute, as seems likely, this will imply that the ground state cannot be eigenstate of $\phi_T$.  Rather, it will be in a wave function of $\phi_T$ eigenvalues, schematically $e^{i\epsilon\phi_T}$.  This wave function must be single-valued as $\phi_T$ is periodic, implying that the oscillon's amplitude $\epsilon$ is quantized.  This is not surprising, as the Sine-Gordon breather, while not an oscillon, satisfies all of the conditions that we have placed upon our theories and solutions and so should be included in our analysis, and its spectrum is known to be quantized \cite{dhnsg}.

\section{Conclusions}

In contrast to current belief, we showed that quantum oscillons do exist. In the leading order expansion in the amplitude  $\epsilon$ and coupling constant $g$, oscillons do not radiate and indeed go back to themselves after one period. Of course, like their classical counterparts, quantum oscillons decay, but this occurs very slowly, allowing oscillons to actively participate in quantum processes. 

Our construction was in 1+1 dimensions, whereas the oscillons which are relevant in astrophysics are (3+1)-dimensional solutions.  Should one expect our conclusion to nonetheless apply beyond 1+1 dimensions?  The key step in our derivation is the fact that the states that decay quickly via the mechanism of Ref.~\cite{hertz} are unsqueezed coherent states, but that squeezed coherent states have a much lower energy.  This difference not only persists beyond 1+1 dimensions, but in fact becomes greater as Coleman has shown \cite{erice} that beyond 1+1 dimensions, unsqueezed coherent states have an infinite energy density while the squeeze was shown in \cite{me4dlett} to reduce this energy density down to a finite quantity.  Furthermore, the squeeze beyond 1+1 dimensions continues to be that which puts each normal mode in its ground state, implying that it prevents radiation at the linear level (\ref{ut}). 

This result opens novel, previously not considered possibilities for the applications of oscillons. Definitely, they are relevant even in the quantum regime of both cosmological and fundamental theories. 

One scenario is that quantum oscillons may dominate the particle spectrum of cosmological models. This concerns models with a real scalar field (inflaton) but also models with a complex or more complicated target space (e.g., the axion). 

Another option is their relevance as dark matter. One seriously considered dark matter candidate is the $Q$-ball \cite{kusenko}, a non-topological soliton stabilized by a Noether charge arising from a $U(1)$ global symmetry. However, it is known that models that support $Q$-balls also support oscillons \cite{Qball-osc}. Hence, even though they are ultimately unstable, quantum oscillons may significantly change the particle content of $Q$-ball theories. This can obviously affect the abundance and properties of the resulting dark matter.  

Furthermore, one should search more extensively for the electroweak oscillon \cite{Gr} as a long-lived non-perturbative excitation, which may appear via decay of the sphaleron \cite{KM}, as often  happens in less complicated models \cite{MR}.

Quantum oscillons may also be relevant in low energy QCD, where other types of solitons, Skyrmions \cite{Sk} and Hopfions \cite{FN}, after quantization, are known to describe baryons \cite{ANW} (atomic nuclei \cite{Mant,Sk1,Sk2,Sk3,Sk4,Sk5,Sk6}) and glueballs \cite{FNW, Chi} respectively. In fact, some long-lived oscillating pionic configurations exist in the chiral Lagrangian up to order $O(p^2)$ \cite{QCD}. The phenomenological imprint of oscillons in this context remains to be discovered. 

In the future we will extend our analysis to higher orders to $\epsilon$ and $g$ to determine the dominant oscillon decay mode in the vacuum, and when subjected to incident radiation \cite{yam}.  In Eq.~(\ref{eveq}) we saw that, at the leading order in $g$, the fields obey the classical equations of motion.  Classically there is no radiation at any order in $\epsilon$, and so we expect that the same will be true in quantum field theory.  The leading decay will then arise from perturbative corrections in $g$, corresponding to multi-loop diagrams.
\def\appendix{\setcounter{section}{0}
 \def\thesection{Appendix \Alph{section}}
 \def\thesubsection{\Alph{section}.\arabic{subsection}}
 \def\theequation{\Alph{section}.\arabic{equation}}}

\appendix
\section{Collective Coordinates} \label{ccapp}

The discrete operators $\phi_0$, $\hat\epsilon$, $\phi_T$ and $\pi_0$ that appeared in our decomposition (\ref{decd}) are not collective coordinates, but they are proportional to collective coordinates up to nonlinear corrections which are suppressed by powers of the coupling $g$.  We will now describe their relationship.

Consider a family of solutions $f_\epsilon(\epsilon x,t)$ of the classical equations of motion, indexed by a parameter $\epsilon$.  Translation and boost invariance, as well as shifts in the parameter, lead to a four-parameter family of solutions
\bea
\phi(x,t)&=&f_{\epsilon+\delta}\left(\gamma(\epsilon+\delta)(x-x_0-v (t-t_0)),\right. \label{phi4}\\
&&\left.\gamma(t-t_0-v(x-x_0))\right)\hsp \gamma=\frac{1}{\sqrt{1-v^2}} \nonumber
\eea
where $x_0,\ t_0,\ \delta$\ and $v$ are collective coordinates.

We wish to expand these solutions about the fiducial solution
\beq
x_0=t_0=v=\delta=0.
\eeq
Let us name first derivatives of (\ref{phi4}) with respect to our collective coordinates, evaluated at this fiducial solution
\bea
\g_B(x,t)\,\,&=&\,\,\frac{1}{c_B}\partial_{x_0}\phi(x,t)\,=\,-\frac{1}{c_B}\partial_x f_{\epsilon}(\epsilon x,t)\\ \g_{T}(x,t)\,&=&\,\frac{1}{c_B}\partial_{t_0}\phi(x,t)=-\frac{1}{c_B}\partial_t f_{\epsilon}(\epsilon x,t)\nonumber\\
\g_M(x,t)&=&t\g_B(x,t)+x\g_T(x,t)=\frac{1}{c_B} \partial_{v}\phi(x,t)\nonumber\\
\g_{\epsilon}(x,t)&=&\frac{1}{c_\epsilon} \partial_{\delta}\phi(x,t)=\frac{1}{c_\epsilon} \partial_\epsilon f_\epsilon(\epsilon x,t)
\nonumber
\eea
where the constants of proportionality $c_B$,  and $c_\epsilon$ can be chosen to simplify computations.
The leading order difference between the solutions (\ref{phi4}) and $f_{\epsilon}(\epsilon x,t)$ is then
\bea
\phi(x,t)-f_{\epsilon}(\epsilon x,t)&\approx&c_B x_0\g_{B}(x,t)+c_B t_0\g_{T}(x,t)\\
&&+c_B v\g_{M}(x,t)+c_\epsilon \delta\g_{\epsilon}(x,t)\nonumber\\
&=&c_B \left(x_0+vt
\right)\g_{B}(x,t)\nonumber\\
&&+c_B (t_0+vx)\g_{T}(x,t)+c_\epsilon\delta \g_\epsilon(x,t).\nonumber
\eea
Here the $\approx$ symbol means equality up to terms quadratic in the collective coordinates.  The first time derivative, again up to corrections quadratic in the collective coordinates, is
\bea
\pi(x,t)-\partial_tf_{\epsilon}(\epsilon x,t)&\approx& c_B \left(x_0+vt
\right)\dot\g_B(x,t)\\
&&+c_B (t_0+vx)\dot\g_{T}(x,t)\nonumber \\
&& +c_B v\g_{B}(x,t)+c_\epsilon \delta \dot \g_\epsilon(x,t).\nonumber
\eea

We will restrict our attention to the special case in which $f_{\epsilon}(\epsilon x,t)$ is periodic with period $2\pi/\Omega$.  This includes quasibreathers
.  We are motivated by small amplitude $\epsilon$ oscillons, which are equal to quasibreathers to all orders in $\epsilon$ and so, in the context of an $\epsilon$ expansion, we expect the results below to apply to oscillons as well.

Let us now restrict our attention to solutions $f_\epsilon$ which are periodic perturbations of (\ref{phi4})
\bea
\phi(x,t)&=&f_{\epsilon+\delta}\left(\gamma(\epsilon+\delta)(x-x_0-v (t-t_0))\right.\\
&&\left.,\gamma(t-t_0-v(x-x_0))\right)\nonumber\\
&&+\pin{k}\left(\g^*_{-k}(x,t) b_{-k}  + \g_{k}(x,t)b^*_k \right)+O(\g_k^2)\nonumber\\
\g_k(x,t)&=&e^{2\pi i\ok{}/\Omega}\g_k\left(x,t+\frac{2\pi}{\Omega}\right),\ \  \ok{}=\sqrt{m^2+k^2}
\nonumber
\eea
where $m$ is the mass of the scalar.  The Floquet functions $\g_k(x,t)$ will play the role played by normal modes in the case of a time-independent solution $f$.  The order $O(\g^2)$ corrections are necessary for $\phi(x,t)$ to solve the equations of motion.

The difference between this perturbed solution and our fiducial solution is
\bea
\phi(x,t)-f_{\epsilon}(\epsilon x,t)&\approx&c_B \left(x_0+vt
\right)\g_{B}(x,t)\label{deccc}\\
&&\hspace{-.5cm}+c_B (t_0+vx)g_{T}(x,t)+c_\epsilon \delta \g_\epsilon(x,t)\nonumber\\
&&\hspace{-.5cm}+\pin{k}\left(\g^*_{-k}(x,t) b_{-k}  + \g_{k}(x,t)b^*_k \right)\nonumber\\
\pi(x,t)-\partial_tf_{\epsilon}(\epsilon x,t)&\approx&c_B \left(x_0+vt
\right)\dot\g_{B}(x,t)\nonumber\\
&&\hspace{-2.5cm}+c_B (t_0+vx)\dot\g_{T}(x,t)+c_B v\g_{B}(x,t))+c_\epsilon \delta \dot \g_\epsilon(x,t)\nonumber\\
&&\hspace{-2.5cm}+\pin{k}\left(\dot{\g}^*_{-k}(x,t) b_{-k} + \dot{\g}_{k}(x,t)b^*_k\right).\nonumber
\eea
Here the $\approx$ symbol now means equality up to terms quadratic in the collective coordinates and the coefficients $b_k$.

At $t=0$, we choose the index $k$ so that
\beq
\g_k^*(x,0)=\g_{-k}(x,0)\hsp \dot{\g}_k^*(x,0)=-\dot{\g}_{-k}(x,0).
\eeq

Not only is the expression for the field in terms of the collective coordinates complicated, but also in the quantum theory the commutation relations obeyed by the collective coordinates are quite complicated.  As a result, collective coordinates are not a convenient choice for quantization.  

It is for this reason that, in the body of the paper, we use a different expansion (\ref{decd}) in terms of $\phi_0$, $\phi_T$, $\hat{\epsilon}$\ and $\pi_0$.  This linear expansion is exact, with no constant offset and no nonlinear corrections.  These four quantities are, at leading order in the coupling, proportional to the collective coordinates with constants of proportionality $c_B$ and $c_\epsilon$.  However, beyond this order, the relationship between these four quantities and the collective coordinates is nonlinear and includes a dependence on the Floquet coefficients $b_k$ and $b^*_k$.  In summary, our approach is to replace the nonlinearity in the decomposition (\ref{deccc}) of $\phi(x)$ and $\pi(x)$ with a nonlinearity in the relationship between our coefficients and the collective coordinates.

\section{Floquet Modes and Their Orthogonality} \label{flapp}

Define the matrix
\beq
M=\left(
\begin{tabular}{ll}
$\partial^2_{\epsilon x}-1+4\sech^2(\epsilon x)\ \ $&$2\sech^2(\epsilon x)$\\
$2\sech^2(\epsilon x)$&$\partial^2_{\epsilon x}-1+4\sech^2(\epsilon x)$
\end{tabular}
\right).
\eeq
Then, using the classical equations of motion, the Floquet modes (\ref{g1eq}) satisfy the coupled equations
\beq
M\left(
\begin{tabular}{l}
$\cG_k(\epsilon x)$ \\
$\cH_k(\epsilon x)$ 
\end{tabular}
\right)=\frac{2m\ok{}}{\epsilon^2}\left(
\begin{tabular}{l}
$-\cG_k(\epsilon x)$ \\
$\cH_k(\epsilon x)$ 
\end{tabular}
\right).
\eeq
Only if $\ok{}$ is much larger than $O(\epsilon^2)$ then the solutions will be plane waves. 

The solutions to these equations include not only the continuum modes in Eq.~(\ref{fl}), but also the zero modes $\g_B$ and $\g_T$ which correspond to space and time translation respectively.  However they do not include $\g_M$ and $\g_\epsilon$.  This is because $\g_M$ corresponds to a boost, and after one period a boost leads to a translation, not a phase.  Similarly, $\g_\epsilon$ corresponds to a change in the amplitude, which changes the period itself, and so leads to a secular evolution over each of the old periods, which again is not a phase.

The Floquet modes satisfy a kind of orthogonality relation.  To see this, compute 
\beq
\int dx \left( \cG_{-k_1}(\epsilon x) \cH_{-k_1}(\epsilon x)\right)M
\left(
\begin{tabular}{l}
$\cG_{k_2}(\epsilon x)$\\
$\cH_{k_2}(\epsilon x)$
\end{tabular}
\right)
\eeq
by acting $M$ to the right, to obtain
\beq
\frac{2m\ok{2}}{\epsilon^2}\int dx \left(-\cG_{-k_1}(\epsilon x)\cG_{k_2}(\epsilon x)+\cH_{-k_1}(\epsilon x)\cH_{k_2}(\epsilon x)\right)
\eeq
and set it equal to the value one would get instead acting $M$ to the left
\beq
\frac{2m\ok{1}}{\epsilon^2}\int dx \left(-\cG_{-k_1}(\epsilon x)\cG_{k_2}(\epsilon x)+\cH_{-k_1}(\epsilon x)\cH_{k_2}(\epsilon x)\right).
\eeq
One finds that if $\ok{1}\neq\ok{2}$ then these integrals vanish
\beq
\int\, dx \left(\cG_{-k_1}(\epsilon x)\cG_{k_2}(\epsilon x)\,-\,\cH_{-k_1}(\epsilon x)\cH_{k_2}(\epsilon x)\right)\,=\,C_{k_1} \delta(k_1-k_2).
\eeq
The case in which they are equal is easily handled, using the explicit forms of the Floquet modes, to obtain
\beq
C_{k}=\frac{4\pi m^2\ok{}^2}{\epsilon^2\sqrt{2m\ok{}-\epsilon^2}} \label{ck}
\eeq
and to show that the integrals vanish when $k_1=-k_2$.

Similarly, evaluating
\beq
\int dx \left( \cH_{-k_1}(\epsilon x) \cG_{-k_1}(\epsilon x)\right)M
\left(
\begin{tabular}{l}
$\cG_{k_2}(\epsilon x)$\\
$\cH_{k_2}(\epsilon x)$
\end{tabular}
\right)
\eeq
by acting $M$ on both sides, one obtains the second orthogonality relation
\beq
\int\, dx \left(\cG_{-k_1}(\epsilon x)\cH_{k_2}(\epsilon x)-\cH_{-k_1}(\epsilon x)\cG_{k_2}(\epsilon x)\right)=0
\eeq
which now applies unless $\ok{1}=-\ok{2}$.  In other words, it applies to all Floquet modes except for the zero modes $\g_B$ and $\g_T$.

With these two orthogonality relations in hand, we may define the projection
\bea
I_k&=&\int dx \left[ \left(\cG_{-k}(\epsilon x)-\cH_{-k}(\epsilon x)\right)\phi(x)\right.\\
&&\left.-\frac{i}{\Omega} \left(\cG_{-k}(\epsilon x)+\cH_{-k}(\epsilon x)\right)\pi(x)
\right]\nonumber
\eea
where $k$ may be a real number, representing a continuum mode, or a discrete index corresponding to one of the zero modes $\g_B$ and $\g_T$.  The orthogonality relations imply that the integral projects out all Floquet modes except for $k$, and so the $I_k$ will in general contain the coefficient of $\cG_k$.  The projector does not eliminate the two non-Floquet modes $\g_\epsilon$ and $\g_M$, and so $I_k$ will generally be a superposition of the coefficients $\pi_0$ and $\hat{\epsilon}$ of $\g_M$, $\g_\epsilon$ as well as the $b_k$  operator that is the coefficient of $\g_k$.

In the case of zero modes, the coefficient $C_k$ in Eq.~(\ref{ck}) vanishes and so the projector eliminates the zero modes themselves, leaving only the coefficients $\pi_0$ and $\hat{\epsilon}$.  The result is the first two equations in Eq.~(\ref{pi0}).  In the case of the continuum modes, there are contributions from the continuum mode itself as well as $\pi_0$ and $\hat{\epsilon}$
\beq
I_k= \frac{C_k}{\pi}b_k+\alpha_k\hat{\epsilon}+\beta_k\pi_0
\eeq
where
\beq
\alpha_k=-\frac{\pi\ok{}^2}{m^2\epsilon}\ \sech\left(\frac{k\pi}{2m}\right)
\hsp
\beta_k=\frac{\pi k \ok{}^2}{3m^2\epsilon \Omega}\sech\left(\frac{k\pi}{2m}\right).
\eeq
Subtracting the $\alpha_k$ and $\beta_k$ contributions, one arrives at the last line of Eq.~(\ref{pi0}).

These expressions for $\pi_0$, $\hat\epsilon$ and $b_k$  allow us to use the canonical commutation relations satisfied by $\phi(x)$ and $\pi(x)$ to demonstrate that $\pi_0$, $\hat\epsilon$ and $b_k$ all commute.  This is essential because our oscillon state $\vac_0$ is a simultaneous eigenstate of these operators, and so it only exists if they commute.

We expect that this same procedure may be used to construct the ground state of the Q-ball.

\section{The Evolution of the Operators} \label{evapp}
At timescales of order $O(\g^0)$, the time evolution of the ket $\vac_0$ is generated by $H\p_2$.  Integrating this evolution over time, one arrives at the evolution operator $U(t)$ defined in Eq.~(\ref{ut}).

Let us define
\beq
\phi_t(x)=U^\dag(t)\phi(x)U (t)\hsp
\pi_t(x)=U^\dag(t)\pi(x)U (t).
\eeq
These would be the field and its conjugate momentum in the interaction picture, but we will continue to work in the Schrodinger picture where these are interpreted as families of operators indexed by the real number $t$.  Using the identity
\beq
\partial_t U(t)=-iU(t) H\p_2(t)
\eeq
one finds
\bea
\partial_t \phi_t(x)&=&iU^\dag(t)[H\p_2(t),\phi(x)]U(t)\label{dt}\\
&=&U^\dag(t) \pi(x) U(t)=\pi_t(x)\nonumber\\
\partial_t \pi_t(x)&=&iU^\dag(t)[H\p_2(t),\pi(x)]U(t)\nonumber\\
&=&i U^\dag(t) \left(\partial_x^2-V^{(2)}(gf(x,t)) \right)\phi(x) U(t)\nonumber\\
&=&\left(\partial_x^2-V^{(2)}(gf(x,t)) \right)\phi_t(x).\nonumber
\eea
Together with the initial conditions
\beq
\phi_0(x)=\phi(x)\hsp \pi_0(x)=\pi(x)
\eeq
Eq.~(\ref{dt}) completely determines the operators $\phi_t(x)$ and $\pi_t(x)$.  Using the decomposition (\ref{decd}) one then arrives the solution (\ref{pev}).

\section* {Acknowledgement}

\noindent
JE was supported by the Higher Education and Science Committee of the Republic of Armenia (Research Project No. 24RL-1C047). K. S. acknowledges financial support from
the Polish National Science Center (Grant No. NCN 2021/43/D/ST2/01122). K. S. and A. W. acknowledge support from the Spanish Ministerio de Ciencia e Innovacion
(MCIN) with funding from the European Union NextGenerationEU (Grant No. PRTRC17.I1) and the Consejeria de Educacion from JCyL through the QCAYLE project, as well as the grant PID2023-148409NB-I00 MTM.


\end{document}